\documentclass{Interspeech}

\usepackage[table,xcdraw]{xcolor}
\usepackage[printonlyused]{acronym} 
\usepackage{pifont}
\usepackage{setspace}

\acrodef{BM}  {background music}
\acrodef{MS}  {multispeaker}
\acrodef{FL}  {foreign languages}
\acrodef{MT}  {multitask}
\acrodef{MTL}  {multitask learning}
\acrodef{ST}  {single-task}
\acrodef{AITW} {Annotated in-the-wild}
\acrodef{LR}  {logistic regression}

\acrodef{AC}    {acoustic condition}
\acrodef{AIO}   {all-in-one}
\acrodef{ANN}   {artificial neural network}
\acrodef{ASR}   {automatic speech recognition}
\acrodef{BCE}   {binary cross-entropy}
\acrodef{BLSTM} {bidirectional long short term memory}
\acrodef{BS}    {batch size}
\acrodef{BSS}   {blind source separation}
\acrodef{CB}    {convolutional block}
\acrodef{CD}    {convolutional decoder}
\acrodef{CE}    {convolutional encoder}
\acrodef{CN}    {channel-wise layer normalization}
\acrodef{CSM}   {clean speech mixtures}
\acrodef{cLN}   {cummulative layer normalization}
\acrodef{Conformer}{convolution augmented Transformer}
\acrodef{Conv-TasNet}{convolutional time-domain audio separation network}
\acrodef{ConSepT}{convolutional separation Transformer}
\acrodef{CP-WER}{concatenated minimum-power word error rate}
\acrodef{CTC}   {connectionist temporal classification}
\acrodef{CV}    {computer vision}
\acrodef{DAE}   {denoising autoencoder}
\acrodef{DANet} {deep attractor network}
\acrodef{DC}    {deep clustering}
\acrodef{DE}    {density estimation}
\acrodef{DL}    {deep learning}
\acrodef{DM}    {dynamic mixing}
\acrodef{DNN}   {deep neural network}
\acrodef{DPRNN} {dual path recurrent neural network model}
\acrodef{DPTNet}{dual path Transformer network}
\acrodef{D-Conv}{depthwise convolution}
\acrodef{DD-Conv}{deformable depthwise convolution}
\acrodef{DDS-Conv}{deformable depthwise-separable convolution}
\acrodef{DP}    {dual-path}
\acrodef{DS-Conv}{depthwise separable convolution}
\acrodef{DTCN}  {deformable temporal convolutional network}
\acrodef{E2E}   {end-to-end}
\acrodef{ED}    {epoch duration}
\acrodef{EER}   {equal error rate}
\acrodef{ER}    {early reflection}
\acrodef{ESTOI} {extended short-time objective intelligibility}
\acrodef{FLOPS} {floating operations per second}
\acrodef{FNR}   {false negative rate}
\acrodef{FPR}   {false positive rate}
\acrodef{FT}    {fine-tuning}
\acrodef{FFT}   {fast Fourier transform}
\acrodef{GFLOPS} {giga floating operations per second}
\acrodef{gLN}   {global layer normalization}
\acrodef{GN}    {group normalization}
\acrodef{GLU}   {gated linear unit}
\acrodef{GRU}   {gated recurrent unit}
\acrodef{GPIT}  {guided permutation invariant training}
\acrodef{GPU}   {graphics processing unit}
\acrodef{GUI}   {graphical user interface}
\acrodef{ICA}   {independent component analysis}
\acrodef{ITW}{in-the-wild}
\acrodef{IN}    {instance normalization}
\acrodef{IPD}   {inter-phase difference}
\acrodef{IR}    {impulse response}
\acrodef{L} {large}
\acrodef{LD}    {logit difference}
\acrodef{LGRU}  {light gated recurrent unit}
\acrodef{LN}    {layer normalization}
\acrodef{LSTM}  {long short term memory}
\acrodef{LTI}   {linear time-invariant}
\acrodef{LUFS}   {loudness units relative to full scale}
\acrodef{M} {medium}
\acrodef{MACs}  {mutiply-accumulate operations}
\acrodef{MHA}   {multihead attention}
\acrodef{MHSA}  {multihead self-attention}
\acrodef{MHCA}  {multihead cross-attention}
\acrodef{MIMO}  {multiple-input multiple-output}
\acrodef{ML}    {machine learning}
\acrodef{MLP}   {multi-layer perceptron}
\acrodef{MOS}   {mean opinion score}
\acrodef{MPGT}  {multi-phase gammatone}
\acrodef{MR}    {mask refinement}
\acrodef{MRD}   {mask refinement decoder}
\acrodef{MSE}   {mean square error}
\acrodef{MVDR}  {minimum-variance distortionless response}
\acrodef{NN}    {neural network}
\acrodef{NMF}   {non-negative matrix factorization}
\acrodef{NSM}   {noisy speech mixture}
\acrodef{NRSM}  {noisy reverberant speech mixture}
\acrodef{OOM}   {out of memory}
\acrodef{ORC-WER}{optimal reference combination word error rate}
\acrodef{PCA}   {principal component analysis}
\acrodef{PE}    {positional encoding}
\acrodef{PESQ}  {perceptual evaluation of speech quality}
\acrodef{PIT}   {permutation invariant training}
\acrodef{PM}    {post-masking}
\acrodef{PMD}   {post-masking decoder}
\acrodef{PReLU} {parametric rectified linear unit}
\acrodef{P-Conv}{\texorpdfstring{pointwise convolution}{P-Conv}}
\acrodef{POLQA} {perceptual objective listening quality assessment}
\acrodef{QDPN}  {quasi-dual-path network}
\acrodef{QRNN}  {quasi-recurrent neural network}
\acrodef{ReLU}  {rectified linear unit}
\acrodef{RF}    {receptive field}
\acrodef{RIR}   {room impulse response}
\acrodef{RNN}   {recurrent neural network}
\acrodef{RSM}   {reverberant speech mixture}
\acrodef{RTF}   {real time factor}
\acrodef{RQ}    {research question}
\acrodef{S} {small}
\acrodef{SA}    {self-attention}
\acrodef{SAD}   {self-attention decoder}
\acrodef{SAE}   {self-attention encoder}
\acrodef{SC}    {skip connection}
\acrodef{SE}    {squeeze-and-excite}
\acrodef{SepFormer}{separation Transformer}
\acrodef{SISDR} {scale-invariant signal-to-distortion ratio}
\acrodef{SRMR}  {speech-to-reverberation modulation energy ratio}
\acrodef{SDR}   {signal-to-distortion ratio}
\acrodef{SiLU}  {sigmoid linear unit}
\acrodef{SISO}  {single-input single-output}
\acrodef{SNMF}   {sparse non-negative matrix factorization}
\acrodef{SNR}   {signal-to-noise ratio}
\acrodef{SOT}   {serialized output training}
\acrodef{SOTA}  {state-of-the-art}
\acrodef{SP}    {signal processing}
\acrodef{SSR}   {speech-to-speech ratio}
\acrodef{SSSR}  {self-supervised speech representation}
\acrodef{STFT}  {short-time Fourier transform}
\acrodef{STOI}  {short-time objective intelligibility}
\acrodef{SVCCA} {singular vector canonical correlation analysis}
\acrodef{SW}    {shared weights}
\acrodef{TAN}   {transformer-attention network}
\acrodef{TasNet}{time-domain audio separation network}
\acrodef{TC}    {time-complexity}
\acrodef{TD-Conformer}{time domain Conformer}
\acrodef{TCN}   {temporal convolutional network}
\acrodef{TF}    {time-frequency}
\acrodef{TPU}   {tensor processing unit}
\acrodef{TSE}   {target speaker extraction}
\acrodef{TSL}   {training signal length}
\acrodef{TTS}   {text to speech}
\acrodef{uPIT}  {utterance-level permutation invariant training}
\acrodef{UPGMA} {unweighted pair group method with arithmetic mean}
\acrodef{WD-Conv}{weighted multi-dilation depthwise convolution}
\acrodef{WD-TCN}{weighted multi-dilation temporal convolutional network}
\acrodef{WE}    {Whisper encoder}
\acrodef{WER}   {word error rate}
\acrodef{WPE}   {weighted prediction error}
\acrodef{XL} {extra-large}

\newcommand{\cmark}{\ding{51}}%
\newcommand{\xmark}{\ding{55}}%

\newcommand{\vek}[1]{\ensuremath{\mathbf{#1}}}    

\newcommand{\Real}{\mathbb{R}}

\interspeechcameraready 

\title{Whilter: A Whisper-based Data Filter for ``In-the-Wild" Speech Corpora Using Utterance-level Multi-Task Classification
\texorpdfstring{\thanks{*Joint contribution.}}{}}

\author[affiliation={}]{William}{Ravenscroft*}
\author[affiliation={}]{George}{Close*}
\author[affiliation={}]{Kit}{Bower-Morris}
\author[affiliation={}]{\texorpdfstring{\par Jamie}{Jamie}}{Stacey}
\author[affiliation={}]{Dmitry}{Sityaev}
\author[affiliation={}]{Kris}{Y. Hong}

\affiliation{}{ConnexAI}{United Kingdom}

\email{\{william.ravenscroft,george.close,kit.bower-morris\}@connex.ai\\\{jamie.stacey,dmitry.sityaev,kris.hong\}@connex.ai}

\keywords{data filtering, speech processing, whisper, multi-task, classification}

\usepackage{comment}

\begin{document}

\maketitle

\begin{abstract}
    Large-scale in-the-wild speech datasets have become more prevalent in recent years due to increased interest in models that can learn useful features from unlabelled data for tasks such as
    speech recognition or synthesis. These datasets often contain undesirable features, such as multiple speakers, non-target languages, and music, which may impact model learning. The Whilter model is proposed as a multitask solution to identify these undesirable samples.
    Whilter uses a Whisper encoder with an attention-based classifier to solve five diverse classification problems at once.
    In addition, an annotated dataset is published for a subset of two popular in-the-wild corpora.
    Whilter achieves F1 scores above $85\%$ and equal error rates of $6.5\%$ to $7.8\%$ for three of five subtasks, outperforming a state-of-the-art BEATs classifier on speech-specific classes, with a notable decrease in processing time compared to a combination of single-task alternatives.
\end{abstract}

\section{Introduction}
Increasingly large corpora have been proposed in recent years from in-the-wild data sources for training deep learning speech processing models \cite{voxceleb2,yodas,emilia}. \Acf{ITW} refers to data that is \textit{not} recorded or collected in a highly controlled environment \cite{voxceleb2}. Rather, \ac{ITW} speech datasets typically consist of audio sourced from public video-sharing platforms and audio podcast aggregators. 
Although \ac{ITW} datasets have been fundamental
in pushing the \ac{SOTA} in speech processing tasks \cite{speartts,mhubert147}, they typically contain potentially undesirable properties such as multi-speaker segments, synthetic speech, non-target languages, background music, and noise.
It is often necessary to identify and \emph{filter} out these samples before model training, e.g. removing Mandarin speech data from English \ac{TTS} model training or vice versa \cite{mandarintts}.

To the best of the authors' knowledge, the specific problem formulation and associated dataset in this paper, motivated by attaining clean \ac{TTS} training data, is novel and has no existing all-in-one solution beyond creating more time-consuming and expensive pipelines of multiple models built for each specific subtask.
Similar solutions have been proposed for other data filtering subtasks, such as transcription quality for \ac{ASR} training data \cite{Georgescu21} or speech / non-speech segmentation \cite{inaspeechsegmenter,mhubert147}. Further, human-annotated labels of subsets of popular \ac{ITW} datasets have been collected and are released as a part of this work along with a Label Studio \cite{labelstudio} \ac{GUI} to enable other researchers to add to the proposed \ac{AITW} dataset themselves\footnote{Link to \ac{AITW} repository: \url{{https://doi.org/10.5281/zenodo.15534661}}}.
\Ac{ST} solutions to the subtasks addressed in this work, such as diarization tools for speaker counting and multispeaker detection \cite{pyannote}, anti-spoofing classification models for synthetic speech detection \cite{for,aasist}, and language classification for non-target language detection \cite{whisper}, have been explored and shown to be solvable using \ac{ML}. However, no \ac{MT} approaches currently exist for the problems outlined here, likely due to the complexity of combining general audio and speech-specific classification tasks. 

In this work, Whilter is proposed for data filtering of \ac{ITW} speech for English TTS training data. Whilter is required to identify multi-speaker audio, the presence of non-English speech (referred to herein as foreign languages), background music, noisy speech (including reverberant speech, i.e. convolutive noise) and synthetic speech. This identifies data to be cleaned before post-processing (e.g. via speech enhancement \cite{pandp} or source separation \cite{tffts}) or discarded entirely if there is potentially corrupting information contained within them (i.e.~synthetic speech and non-target languages in many cases). 
The motivation for using the \ac{MT} approach is threefold. Firstly, due to the size of most \ac{ITW} datasets, it is not viable in terms of time and compute to filter for each class individually, let alone train models for each class. Secondly, the proposed \ac{MT} system allows for selective filtering based on the downstream task, in a `one and done' paradigm. 
Finally, a single network jointly learning all labels might learn to exploit the relationships between speech properties \cite{whisper}. 
The Whisper encoder \cite{whisper} is chosen as an audio feature extractor for Whilter due to recent findings demonstrating the power of speech foundation models for speech quality assessment \cite{pandp, whispersi} and synthetic speech detection \cite{asvspoofbad}.


\section{Whilter}\label{sec:whilter}
\begin{figure}[!ht]
    \centering
    \includegraphics[width=\columnwidth,trim={0.7cm 0cm 0.0cm 2.2cm},clip]{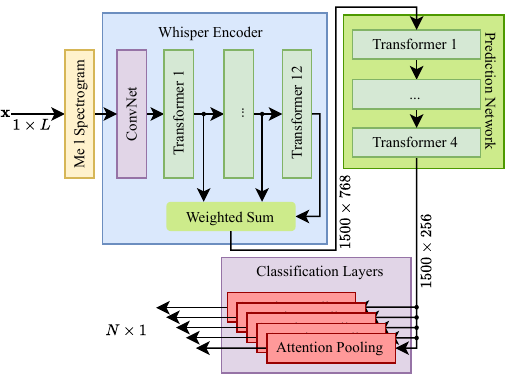}
    \caption{Diagram of the Whilter model, composed of a frozen Whisper encoder with learnable layer weights, a transformer prediction network and attention-pooling-based classification layers.}
    \label{fig:model_schematic}
\end{figure}

The Whilter model $\mathcal{W}$, takes a single channel audio signal of length $L$, denoted $\vek{x}\in\Real^{L}$, and maps it to $N$ output classes, i.e.~$\mathcal{W}:1 \times L \mapsto N \times 1$.
Whilter is composed of three main components: a frozen Whisper \cite{whisper} encoder network, a learnable Transformer \cite{Vaswani} network, and a bank of attention pooling layers \cite{Mittag_2021} for the five output classes shown in \autoref{fig:model_schematic} (multi-speaker audio, background music, foreign language, noisy speech, and synthetic speech).

\subsection{Whisper Encoder}
Motivated by recent work in speech intelligibility and quality assessment \cite{whispersi,sehallucinations}, the encoder part of the \textit{small}  Whisper model\footnote{Link to Whisper models: \url{https://github.com/openai/whisper}} \cite{whisper} is used for extracting rich audio features. Whisper is a multilingual \ac{ASR} foundation model trained using a large-scale weak supervision and \ac{MTL} \cite{mtl} for 
speech translation, spoken language identification, and voice activity detection among other tasks \cite{whisper}. A speech-specific model is chosen over an audio foundation model such as BEATs \cite{beats} or CLAP \cite{clap}, because three of the subtasks are speech-specific (multispeaker, foreign language and synthetic speech classification) and more general audio foundation models may not have a nuanced enough representation of speech. This assumption is validated later in our results, cf.~\autoref{tab:final_table}.
 Whisper inputs are padded or truncated to $30$~s, resulting in a fixed-size of $1500$ encoder output frames with $16$~kHz input audio.
Instead of using just the Whisper encoder output layer directly, the twelve intermediate transformer layer outputs are weighted and summed. The twelve layer weights are learned, while the original Whisper encoder parameters remain frozen, cf.~\autoref{fig:model_schematic}.

\subsection{Transformer Network}\label{sec:transformer}
A $4$ layer self-attentive transformer network \cite{Vaswani} with $4$ heads, denoted $\mathcal{T}$, is trained on top of the Whisper encoder to learn mappings between the Whisper features and output classes.
The network reduces the dimension from $768$ to $256$ at its output, i.e.~$\mathcal{T}:\Real^{1500\times768}\mapsto\Real^{1500\times256}.$

\subsection{Attention Pooling Layers}\label{sec:attpool}
The output from the Transformer layers $\vek{F}\in\Real^{1500\times 256}$ is passed through a linear attention pooling head~\cite{Mittag_2021}, $\mathcal{A}_n $, for each of the $N$ target classes, i.e.~$n\in\{1,\ldots,N\}$. The attention part of the head is composed of a linear layer with a \ac{ReLU} activation, dropout, and another linear layer followed by a softmax operation along the temporal axis. This attention tensor is then matrix-multiplied with the original input \vek{F} along the temporal axis. In the final stage, the feature dimension (of size $256$) is reduced to $1$ by a linear layer, with a residual connection from the mean of the input $\vek{F}$ along the time axis to improve gradient stability. The outputs of each of the $N$ heads are combined into a single tensor $\hat{\vek{y}}\in\Real^{N\times 1}$.

\subsection{Loss Function}
\Acf{BCE} is the loss function used for training the Whilter model. The \ac{BCE} loss between labels $\vek{y}\in\Real^{N\times 1}$ and predictions $\hat{\vek{y}}$ is defined as
\begin{multline}
     \mathcal{L}_{BCE}\left(\hat{\vek{y}},\vek{y}\right)=-\frac{1}{N}\sum_{n=1}^N( \vek{y}_n\log\left(\hat{\vek{y}}_n\right)  \\
     +\left(1-\vek{y}_n\right)\log\left(1-\hat{\vek{y}}_n\right)).
\end{multline}

\section{Data Preparation}\label{sec:data}
A two-stage training approach 
using training with artificially mixed non-\ac{ITW} data, followed by fine-tuning with the \ac{AITW} data.
For the first training stage, simple filename and label pairs are derived from non-\ac{ITW} datasets. Dynamic mixing
 \cite{wavesplit}
, popularized for speech separation \cite{tffts}, is used for augmenting the combinations of these labels, i.e.~mixing music, English speech, noise, foreign languages, and synthetic speech such that each class occurs more frequently in training, accelerating model convergence.
A randomly selected quarter of the batch is artificially mixed with speech. This is also repeated with noise and music samples. All mixing of music, speech and noise happens at \acp{SNR} or \acp{SSR} randomly chosen from a range of $(-5,10)$~dB. Note that this mixing is independent for each class type, i.e.~multispeaker noisy samples containing music do occur.
In both training stages, weighted random sampling was used. This helped to address data imbalances of some class labels, e.g. synthetic speech, and further improve the rate of convergence. To this end, a weight of $\frac{\text{\# negative class labels}}{\text{\# positive class labels}}$ was added to positive labels in each class.

\subsection{Non-ITW Datasets}
Several datasets were selected for the non-\ac{ITW} training stage to diversely represent the target classes, as well as include some data that does not fit into any of the classes. Segmented clips from the AMI \cite{ami} and AliMeeting \cite{alimeeting} corpora were used to represent ``real" multispeaker labels with both overlapped and non-overlapped utterances. For both, the summed headset and single-distant microphone configurations were used. Single-speaker segments were also extracted from these datasets.
Foreign language labels were drawn from subsets of the MLS \cite{mls}, AliMeeting \cite{alimeeting}, FOR \cite{for} and MUSAN \cite{musan} corpora. The FOR \cite{for} and ODSS \cite{odss} corpora were used for synthetic speech labels. 
MUSAN \cite{musan} and OpenMic-2018 \cite{openmic2018} were used for the background music label. 
Noise and noisy speech samples were drawn from MUSAN \cite{musan}, FSDNoisy-18k \cite{fsdnoisy18k}, DEMAND \cite{demand}, AMI \cite{ami} and AliMeeting \cite{alimeeting}.
Finally, some \textit{classless} examples, i.e.~clean, near-field single-speaker English utterances, are derived from Librispeech 960, AMI and MUSAN.


\subsection{\texorpdfstring{\Acf{AITW} Dataset}{Annotated In The Wild (AITW) Datasets}}
Randomly selected subsets of the Emilia \cite{emilia} and YODAS \cite{yodas} datasets were chosen to be manually annotated. 
Data samples were annotated by two experienced speech annotators. Detailed guidelines were created to encourage alignment and accuracy in the resulting dataset. 
\textit{Label Studio} \cite{labelstudio}, a popular open-source data labelling tool, was used for designing a bespoke \ac{GUI} to accelerate the labelling process.
For the multispeaker label, annotators counted the number of unique speakers present in a given audio clip. This was then converted to a boolean multispeaker label (number of speakers $>1$) for Whilter training. The number of speakers is retained in the accompanying dataset.
For the other subtasks, annotators were required to select an integer value, $\{0,1\}=\{\text{False},\text{True}\}$, equating the presence of the class (foreign language, background music, noise, and synthetic speech) in each audio sample.

From the resulting annotated data, three subsets were created for model fine-tuning, validation and testing. 
In total 18,346 samples were annotated for fine-tuning ($\approx 55$~hrs, 1353 for validation ($\approx 4$~hrs) and 1716 for testing ($\approx 5$~hrs), resulting in 21,414 samples in total ($\approx 64$~hrs).
\begin{figure}[!ht]
    \centering
    \includegraphics[]{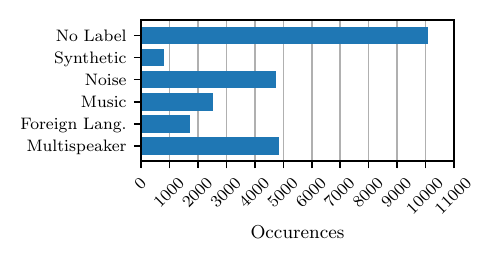}
    \caption{Number of occurrences of each label across the entire \ac{AITW} dataset as well as the total number of samples with no label.}
    \label{fig:aitw_occurrences}
\end{figure}
Label frequency is shown in \autoref{fig:aitw_occurrences}. Notably, the synthetic speech label is the sparsest class label, which may impact model performance. This is discussed more in \autoref{sec:discussion}.


\section{Experimental Setup}\label{sec:expsetup}

\subsection{Training Configurations}
The model was trained using an ADAM optimizer \cite{adam} and an exponentially decaying learning rate scheduler whereby the learning rate, $\eta$, was decayed by a factor, $\gamma$, at the end of each epoch. For both training stages, $15,000$ samples are selected per epoch using a weighted random sampler. A batch size of $64$ is used, resulting in $235$ iterations per epoch.
In the simulated training stage, the model is trained for $10$ epochs with $\eta=10^{-5}$ and $\gamma=0.7$. In the fine-tuning stage, the model is trained for a further $100$ epochs with $\eta=10^{-5}$ and $\gamma=0.98$. 
For the fine-tuning stage, data augmentation techniques are also used in place of the dynamic mixing of the simulated training to prevent model overfitting. The augmentation techniques include frequency dropping, frame dropping, bit-resolution reduction, sign flipping and speed perturbation \cite{dataaug}. Note, none of these techniques were included under the definition of \textit{noise} in the annotation guidelines.

\subsection{Baselines}
The baselines are separated into two classes of model type: \acf{ST} and \acf{MT} models, i.e. models that solve a single subtask or models that solve multiple but not necessarily all subtasks, respectively.

\Ac{ST} Models: A series of open-source \ac{SOTA} models are used for evaluating some of the subtasks in a stand-alone setting. {Pyannote} \cite{pyannote}, a popular speaker diarization model, is used as a baseline for speaker counting and thus multispeaker classification. The small and large-v3 {Whisper} models \cite{whisper} are used as a baseline for foreign language detection, using the language label of the model to differentiate English from non-English.
{DNSMOS} \cite{dnsmos} is a popular speech quality estimator widely used for scoring denoising models in terms of background noise suppression, signal quality and overall quality. To create a baseline for the noise class, a \ac{LR} model was fitted on top of the raw output DNSMOS values using the \ac{AITW} fine-tuning data.
AASIST \cite{aasist}, a popular anti-spoofing model, is used as a baseline for synthetic speech classification. The model is evaluated both before and after being fine-tuned on the \ac{AITW} data.

\Ac{MT} Models:  
The {inaSpeechSegmenter} model \cite{inaspeechsegmenter} has become a popular choice for pre-annotating audio into music, speech and noise chunks \cite{mhubert147}. Thus, in this work, we assess its capabilities for music and noise classification in audio.
The {BEATs} \cite{beats} model is a \ac{SOTA} general audio foundation model for audio classification. It is used in this work as a \ac{MT} baseline fine-tuned on the \ac{AITW} fine-tuning set. BEATs computes a sequence of {768}-dimensional features (the same as the small Whisper model used in Whilter) from a given time-domain audio signal. This is used to validate our choice of the speech-specific foundation model in our choice of input features, as well as compare to a \ac{SOTA} audio classification model. For fairness, we also replace the single linear classification layer of BEATs with the \acf{TAN} used in Whilter, cf.~\autoref{sec:transformer} and \autoref{sec:attpool}.

\subsection{Evaluations Metrics}
Standard classification measures are used for assessing classifier accuracy across each subtask. These include \ac{FPR}, \ac{FNR}, \ac{EER}, Precision, Recall and F1 score \cite{sedstats}. In addition, we also compare and contrast both \acf{ST} and \acf{MT} approaches in terms of their average processing time, denoted $\bar{T}_\mathrm{proc}$, for each sample in the test set on an Nvidia A10 GPU.

\section{Results}\label{sec:results}
\begin{table*}[!tb]
\centering
\setlength{\tabcolsep}{3pt}
\caption{Performance metrics comparing the Whilter model to open source baselines. The mean processing time, $\bar{T}_\mathrm{proc}$, is reported in seconds (s). FT indicates models that have been fine-tuned on the proposed \ac{AITW} dataset. MT indicates model that perform multitask classification. Values in bold indicate the best performance of a given metric for a given class. Model Size is reported in number of parameters (both frozen and unfrozen in training).}
\label{tab:final_table}
\begin{tabular}{|c|c|c|c|ccc|ccc|c|c|}
\hline
\rowcolor[HTML]{C0C0C0} 
\textbf{Class}   & \textbf{Model} & \textbf{MT} & \textbf{FT} & \textbf{FPR \%} & \textbf{FNR \%} & \textbf{EER \%} & \textbf{Prec. \%} & \textbf{Rec. \%} & \textbf{F1 \%} & $\bar{T}_\mathrm{proc.}$ & \textbf{Size} \\ \hline
Multispeaker & Pyannote \cite{pyannote}   & \color{red}\xmark  & \color{red}\xmark  & 8.2 & 20.7& 14.5& 69.2  & 79.3 & 73.9   & 0.479 s  & 8.11M\\
Multispeaker & BEATs \cite{beats} & \cmark  & \cmark  & 12.1& 13.3& 12.7& 62.4  & 86.7 & 72.5   & 0.025 s  & 90.3M\\
Multispeaker & BEATs \cite{beats} + TAN& \cmark  & \cmark  & 5.4 & 19.8& 12.6& 77.5  & 80.2 & 78.8   & 0.028 s & 93.3M \\
Multispeaker & \textit{Whilter}   & \cmark  & \cmark  & \textbf{4.4}& \textbf{9.0}& \textbf{6.7}& \textbf{82.8} & \textbf{91.0}& \textbf{86.7}  & 0.033 s & 91.1M \\ \hline
Background Music & inaSpeechSegmenter \cite{inaspeechsegmenter}& \cmark  & \color{red}\xmark  & \textbf{1.9}& 77.3& 39.6& 65.0  & 22.7 & 33.7   & 0.524 s & 0.79M\\
Background Music & BEATs \cite{beats} & \cmark  & \cmark  & 2.8 & \textbf{7.4}& \textbf{5.1}& \textbf{83.5} & \textbf{92.6}& \textbf{87.8}  & 0.025 s  & 90.3M\\
Background Music & BEATs \cite{beats} + TAN& \cmark  & \cmark  & 3.9 & 9.2 & 6.5 & 78.2  & 90.8 & 84.0   & 0.028 s  & 93.3M \\
Background Music & \textit{Whilter}   & \cmark  & \cmark  & 6.5 & 9.2 & 7.8 & 68.2  & 90.8 & 77.9   & 0.033 s & 91.1M \\ \hline
Foreign Language & Whisper (small) \cite{whisper}& \color{red}\xmark  & \color{red}\xmark  & 3.7 & \textbf{9.2}& \textbf{6.5}& 73.4  & \textbf{90.8}& 81.1   & 0.047 s & 244M \\
Foreign Language & Whisper (large) \cite{whisper}& \color{red}\xmark  & \color{red}\xmark  & 3.6 & 11.0& 7.3 & 73.3  & 89.0 & 80.4   & 0.239 s & 1550M \\
Foreign Language & BEATs \cite{beats} & \cmark  & \cmark  & \textbf{0.0}& 100.0   & 50.00   & 0.00  & 0.00 & 0.00   & 0.025 s & 90.3M \\
Foreign Language & BEATs \cite{beats} + TAN& \cmark  & \cmark  & 0.1 & 99.9& 50.00   & 0.00  & 0.00 & 0.00   & 0.028 s & 93.3M \\
Foreign Language & \textit{Whilter}   & \cmark  & \cmark  & 1.5 & 11.6& \textbf{6.5}& \textbf{86.9} & 88.4 & \textbf{87.7}  & 0.033 s & 91.1M \\ \hline
Noise& inaSpeechSegmenter \cite{inaspeechsegmenter} & \cmark  & \color{red}\xmark  & \textbf{1.1}& 95.1& 48.1& 57.1  & 4.9  & 9.0& 0.524 s & 0.79M \\
Noise& DNSMOS \cite{dnsmos} + LR& \color{red}\xmark  & \cmark  & 8.8 & 61.4& 35.1& 57.9  & 38.6 & 46.3   & 0.453 s & 0.29M \\
Noise& BEATs \cite{beats} & \cmark  & \cmark  & 3.8 & 47.7& 25.7& \textbf{81.4} & 52.3 & 63.7   & 0.025 s & 90.3M \\
Noise& BEATs \cite{beats} + TAN& \cmark  & \cmark  & 6.2 & \textbf{36.4}   & \textbf{21.3}   & 76.2  & \textbf{63.6}& \textbf{69.3}  & 0.028 s & 93.3M \\
Noise& \textit{Whilter}   & \cmark  & \cmark  & 5.4 & 45.7& 25.6& 75.8  & 54.3 & 63.2   & 0.033 s & 91.1M \\ \hline
Synthetic Speech & AASIST \cite{aasist}  & \color{red}\xmark  & \color{red}\xmark  &  23.7& 77.1   & 50.4& 2.0   & 22.9  & 3.6& 0.047 s & 0.3M \\
Synthetic Speech & AASIST \cite{aasist}  & \color{red}\xmark  & \cmark  & 27.0 & \textbf{40.0}   & 33.5& 4.4   & \textbf{60.0}  & 8.2& 0.047 s & 0.3M  \\
Synthetic Speech & BEATs \cite{beats} & \cmark  & \cmark  & 0.0 & 100.0   & 50.0& 0.0   & 0.0  & 0.0& 0.025 s & 90.3M \\
Synthetic Speech & BEATs \cite{beats} + TAN& \cmark  & \cmark  & 0.0 & 100.0   & 50.0& 0.0   & 0.0  & 0.0& 0.028 s & 93.3M \\
Synthetic Speech & \textit{Whilter}   & \cmark  & \cmark  & \textbf{0.5}& {62.9}   & \textbf{31.7}   & \textbf{59.1} & {37.1}& \textbf{45.6}  & 0.033 s & 91.1M \\ \hline
\end{tabular}
\end{table*}
Results for both \ac{ST} and \ac{MT} approaches are shown in \autoref{tab:final_table}.
For \ac{ST} multispeaker classification, the Whilter model outperformed the Pyannote model \cite{pyannote} in \ac{EER} by $7.8\%$ and F1 by $12.8\%$.
Whilter was also faster on average than Pyannote by more than a factor of $10$.
Comparing Whilter to Whisper for \ac{ST} foreign language classification, both have comparable $6.5\%$ \acp{EER}, with Whisper small having slightly better \ac{FNR}, $9.2\%$ vs. $11.6\%$ for Whilter. Interestingly, Whisper small outperformed Whisper large in the foreign language classification task. Notably, the much shallower transformer network (4 layers) compared to both Whisper decoders ($12$ and $32$ layers) results in significant improvements in processing time.
The \ac{ST} DNSMOS \cite{dnsmos} model with a \ac{LR} layer fine-tuned on the \ac{AITW} data was outperformed across all metrics, including mean processing time, by Whilter. 
The base AASIST model \cite{aasist} did not generalize well to the \ac{AITW} dataset. This has been investigated further in \cite{mlaad,asvspoofbad}, the root cause is likely due to a domain mismatch in the AntiSpoof-2019 training data. The fine-tuned AASIST model had a better \ac{FNR} and recall than Whilter but Whilter still outperformed in \ac{EER} and F1 score.


Comparing the \ac{MT} approaches, Whilter performs significantly better on both noise and music classification than inaSpeechSegmenter \cite{inaspeechsegmenter}. This is likely caused by domain mismatch of the inaSpeechSegmenter training data, and because when speech overlaps with music or noise the model classifies the audio as just speech, leading to a higher \ac{FNR} for music and noise.
For three out of five of the subtasks, multispeaker, foreign language and synthetic speech, Whilter outperforms the fine-tuned BEATs \cite{beats} and BEATs + TAN \ac{MT} models. Notably, these are the speech-specific tasks, validating the choice of using the Whisper speech foundation model. Further evidence of this is that the BEATs model seems entirely unable to solve the foreign language and synthetic speech tasks. This is a similar finding to \cite{mlaad} where the researchers showed the strength of using speech foundation models for synthetic speech detection.
In general, both Whilter and BEATs models outperform the \ac{ST} models in terms of average processing time, $\bar{T}_\mathrm{proc.}$.

\section{Discussion}\label{sec:discussion}
This paper presents a first effort towards solving a non-trivial \ac{MT} classification problem. A key finding is that speech-specific foundation models provide a clear advantage in foreign language and synthetic speech detection, while audio foundation models perform slightly better with music and noise. Future work could investigate combining speech and audio foundation models or deriving a general audio foundation model capable of learning more nuanced features relating to speech.

The \ac{AITW} dataset was proposed to solve five \ac{TTS}-related data-filtering subproblems. 
Though highly experienced annotators were used, it is hard to have high confidence in some synthetic speech labels due to the nature of the inverse problem trying to be solved by modern \ac{TTS} and voice conversion systems, i.e. these systems are trying to ``solve" for a high degree of realism \cite{aasist}.
Coupled with this, the synthetic speech label is very sparse meaning the number of positive examples in both train and test sets have a limited amount of diversity, impacting both model generalization and testing reliability. In future, it is hoped that by bootstrapping the Whilter model to find more positively classified synthetic labels and then having annotators manually check the labels, the quality and quantity of these labels can be improved.
Further improvements could be made by using more annotators and majority voting on labels for each sample. This may also be relevant for the noise label, where there is possibly some ambiguity in the level of ``noisiness" for convolutive and stationary noise sources.
Finally, this work did not evaluate the trade-offs between \ac{MT} and \ac{ST} approaches using the same underlying model (i.e.~one Whilter model per sub-task). The \ac{MT} approach might encourage the model to exploit relationships between sub-tasks \cite{whisper} however, there is an argument that separate \ac{ST} models may simplify the overall problems and make it easier for each model to solve.

\section{Conclusion}\label{sec:conclusion}
In this work, the Whilter model was proposed along with the \ac{AITW} dataset, for \ac{MT} data filtering of in-the-wild speech recordings. It was shown that the \ac{AITW} dataset can be used to train classifiers on multispeaker, foreign language, background music, noise and synthetic speech labels. Whilter outperformed numerous widely used open-source models that can be used for these subtasks. When compared to the \ac{SOTA} BEATs audio classification model fine-tuned on the \ac{AITW} data, Whilter outperformed on three out of five subtasks, two of which BEATs was unable to solve, demonstrating the strength of Whisper \cite{whisper} as a foundation model for speech-specific classification tasks.

\bibliographystyle{IEEEtran}
{\footnotesize \bibliography{short}}

\begin{thebibliography}{10}
\providecommand{\url}[1]{#1}
\csname url@samestyle\endcsname
\providecommand{\newblock}{\relax}
\providecommand{\bibinfo}[2]{#2}
\providecommand{\BIBentrySTDinterwordspacing}{\spaceskip=0pt\relax}
\providecommand{\BIBentryALTinterwordstretchfactor}{4}
\providecommand{\BIBentryALTinterwordspacing}{\spaceskip=\fontdimen2\font plus
\BIBentryALTinterwordstretchfactor\fontdimen3\font minus \fontdimen4\font\relax}
\providecommand{\BIBforeignlanguage}[2]{{%
\expandafter\ifx\csname l@#1\endcsname\relax
\typeout{** WARNING: IEEEtran.bst: No hyphenation pattern has been}%
\typeout{** loaded for the language `#1'. Using the pattern for}%
\typeout{** the default language instead.}%
\else
\language=\csname l@#1\endcsname
\fi
#2}}
\providecommand{\BIBdecl}{\relax}
\BIBdecl

\bibitem{voxceleb2}
J.~S. Chung, A.~Nagrani, and A.~Zisserman, ``Voxceleb2: Deep speaker recognition,'' in \emph{Interspeech 2018}, 2018, pp. 1086--1090.

\bibitem{yodas}
X.~Li, S.~Takamichi, T.~Saeki, W.~Chen, S.~Shiota, and S.~Watanabe, ``{YODAS: Youtube-Oriented Dataset for Audio and Speech},'' in \emph{ASRU 2023}, 2023, pp. 1--8.

\bibitem{emilia}
\BIBentryALTinterwordspacing
H.~He, Z.~Shang, C.~Wang, X.~Li, Y.~Gu, H.~Hua, L.~Liu, C.~Yang, J.~Li, P.~Shi, Y.~Wang, K.~Chen, P.~Zhang, and Z.~Wu, ``Emilia: An extensive, multilingual, and diverse speech dataset for large-scale speech generation,'' 2024. [Online]. Available: \url{https://arxiv.org/abs/2407.05361}
\BIBentrySTDinterwordspacing

\bibitem{speartts}
E.~Kharitonov, D.~Vincent, Z.~Borsos, R.~Marinier, S.~Girgin, O.~Pietquin, M.~Sharifi, M.~Tagliasacchi, and N.~Zeghidour, ``Speak, read and prompt: High-fidelity text-to-speech with minimal supervision,'' \emph{Transactions of the Association for Computational Linguistics}, vol.~11, pp. 1703--1718, 2023.

\bibitem{mhubert147}
M.~{Zanon Boito}, V.~Iyer, N.~Lagos, L.~Besacier, and I.~Calapodescu, ``mhubert-147: A compact multilingual hubert model,'' in \emph{Interspeech 2024}, 2024, pp. 3939--3943.

\bibitem{mandarintts}
J.~Li, D.~Sityaev, and J.~Hao, ``Sentence level intelligibility evaluation for mandarin text-to-speech systems using semantically unpredictable sentences,'' in \emph{Interspeech 2007}, 2007, pp. 1350--1353.

\bibitem{Georgescu21}
A.-L. Georgescu, C.~Manolache, D.~Oneaţă, H.~Cucu, and C.~Burileanu, ``Data-filtering methods for self-training of automatic speech recognition systems,'' in \emph{2021 SLT Workshop}, 2021, pp. 1--7.

\bibitem{inaspeechsegmenter}
D.~Doukhan, J.~Carrive, F.~Vallet, A.~Larcher, and S.~Meignier, ``An open-source speaker gender detection framework for monitoring gender equality,'' in \emph{ICASSP 2018}.\hskip 1em plus 0.5em minus 0.4em\relax IEEE, 2018.

\bibitem{labelstudio}
\BIBentryALTinterwordspacing
M.~Tkachenko, M.~Malyuk, A.~Holmanyuk, and N.~Liubimov, ``{Label Studio}: Data labeling software,'' 2020-2024. [Online]. Available: \url{https://github.com/HumanSignal/label-studio}
\BIBentrySTDinterwordspacing

\bibitem{pyannote}
H.~Bredin, ``{pyannote.audio 2.1 speaker diarization pipeline: principle, benchmark, and recipe},'' in \emph{Interspeech 2023}, 2023.

\bibitem{for}
R.~Reimao and V.~Tzerpos, ``{FoR}: A dataset for synthetic speech detection,'' in \emph{2019 International Conference on Speech Technology and Human-Computer Dialogue (SpeD)}, 2019, pp. 1--10.

\bibitem{aasist}
J.-w. Jung, H.-S. Heo, H.~Tak, H.-j. Shim, J.~S. Chung, B.-J. Lee, H.-J. Yu, and N.~Evans, ``{AASIST}: {Audio Anti-Spoofing Using Integrated Spectro-Temporal Graph Attention Networks},'' in \emph{ICASSP 2022}, 2022, pp. 6367--6371.

\bibitem{whisper}
A.~Radford, J.~W. Kim, T.~Xu, G.~Brockman, C.~McLeavey, and I.~Sutskever, ``Robust speech recognition via large-scale weak supervision,'' in \emph{ICML 2023}.\hskip 1em plus 0.5em minus 0.4em\relax JMLR.org, 2023.

\bibitem{pandp}
G.~Close, W.~Ravenscroft, T.~Hain, and S.~Goetze, ``Perceive and predict: Self-supervised speech representation based loss functions for speech enhancement,'' in \emph{ICASSP 2023}, 2023, pp. 1--5.

\bibitem{tffts}
W.~Ravenscroft, G.~Close, S.~Goetze, T.~Hain, M.~Soleymanpour, A.~Chowdhury, and M.~C. Fuhs, ``Transcription-free fine-tuning of speech separation models for noisy and reverberant multi-speaker automatic speech recognition,'' in \emph{Interspeech 2024}, 2024, pp. 4998--5002.

\bibitem{whispersi}
R.~Mogridge, G.~Close, R.~Sutherland, T.~Hain, J.~Barker, S.~Goetze, and A.~Ragni, ``Non-intrusive speech intelligibility prediction for hearing-impaired users using intermediate asr features and human memory models,'' in \emph{ICASSP 2024}, 2024, pp. 306--310.

\bibitem{asvspoofbad}
O.~Chetia~Phukan, G.~Kashyap, A.~B. Buduru, and R.~Sharma, ``Heterogeneity over homogeneity: Investigating multilingual speech pre-trained models for detecting audio deepfake,'' in \emph{NAACL 2024}, Jun. 2024, pp. 2496--2506.

\bibitem{Vaswani}
A.~Vaswani, N.~Shazeer, N.~Parmar, J.~Uszkoreit, L.~Jones, A.~N. Gomez, {\L}.~Kaiser, and I.~Polosukhin, ``{Attention is All you Need},'' in \emph{Advances in Neural Information Processing Systems}, I.~Guyon, U.~V. Luxburg, S.~Bengio, H.~Wallach, R.~Fergus, S.~Vishwanathan, and R.~Garnett, Eds., vol.~30.\hskip 1em plus 0.5em minus 0.4em\relax Curran Associates, Inc., 2017.

\bibitem{Mittag_2021}
G.~Mittag, B.~Naderi, A.~Chehadi, and S.~Möller, ``Nisqa: A deep cnn-self-attention model for multidimensional speech quality prediction with crowdsourced datasets,'' in \emph{Interspeech 2021}.\hskip 1em plus 0.5em minus 0.4em\relax ISCA, Aug. 2021.

\bibitem{sehallucinations}
G.~Close, T.~Hain, and S.~Goetze, ``Hallucination in perceptual metric-driven speech enhancement networks,'' in \emph{EUSIPCO 2024}, 2024, pp. 21--25.

\bibitem{mtl}
\BIBentryALTinterwordspacing
S.~Ruder, ``An overview of multi-task learning in deep neural networks,'' 2017. [Online]. Available: \url{https://arxiv.org/abs/1706.05098}
\BIBentrySTDinterwordspacing

\bibitem{beats}
S.~Chen, Y.~Wu, C.~Wang, S.~Liu, D.~Tompkins, Z.~Chen, W.~Che, X.~Yu, and F.~Wei, ``{BEAT}s: Audio pre-training with acoustic tokenizers,'' in \emph{ICML 2023}, A.~Krause, E.~Brunskill, K.~Cho, B.~Engelhardt, S.~Sabato, and J.~Scarlett, Eds., vol. 202, Jul 2023, pp. 5178--5193.

\bibitem{clap}
Y.~Wu, K.~Chen, T.~Zhang, Y.~Hui, T.~Berg-Kirkpatrick, and S.~Dubnov, ``Large-scale contrastive language-audio pretraining with feature fusion and keyword-to-caption augmentation,'' in \emph{ICASSP 2023}, 2023, pp. 1--5.

\bibitem{wavesplit}
N.~Zeghidour and D.~Grangier, ``Wavesplit: End-to-end speech separation by speaker clustering,'' \emph{IEEE/ACM Trans. Audio, Speech and Lang. Proc.}, vol.~29, p. 2840–2849, Jul. 2021.

\bibitem{ami}
J.~Carletta, S.~Ashby, S.~Bourban, M.~Flynn, M.~Guillemot, T.~Hain, J.~Kadlec, V.~Karaiskos, W.~Kraaij, M.~Kronenthal, G.~Lathoud, M.~Lincoln, A.~Lisowska, I.~McCowan, W.~Post, D.~Reidsma, and P.~Wellner, ``{The AMI Meeting Corpus: A Pre-announcement},'' in \emph{Machine Learning for Multimodal Interaction}, S.~Renals and S.~Bengio, Eds.\hskip 1em plus 0.5em minus 0.4em\relax Berlin, Heidelberg: Springer Berlin Heidelberg, 2006, pp. 28--39.

\bibitem{alimeeting}
F.~Yu, S.~Zhang, Y.~Fu, L.~Xie, S.~Zheng, Z.~Du, W.~Huang, P.~Guo, Z.~Yan, B.~Ma, X.~Xu, and H.~Bu, ``{M2Met}: {The {ICASSP} 2022 Multi-Channel Multi-Party Meeting Transcription Challenge},'' in \emph{ICASSP 2022}, 2022, pp. 6167--6171.

\bibitem{mls}
V.~Pratap, Q.~Xu, A.~Sriram, G.~Synnaeve, and R.~Collobert, ``{MLS}: {A Large-Scale Multilingual Dataset for Speech Research},'' in \emph{Interspeech 2020}, 2020, pp. 2757--2761.

\bibitem{musan}
D.~Snyder, G.~Chen, and D.~Povey, ``{MUSAN}: {A} {M}usic, {S}peech, and {N}oise {C}orpus,'' 2015, arXiv:1510.08484v1.

\bibitem{odss}
A.~Yaroshchuk, C.~Papastergiopoulos, L.~Cuccovillo, P.~Aichroth, K.~Votis, and D.~Tzovaras, ``An open dataset of synthetic speech,'' in \emph{IEEE WIFS 2023}, 2023, pp. 1--6.

\bibitem{openmic2018}
E.~J. Humphrey, S.~Durand, and B.~McFee, ``{OpenMIC-2018: An Open Dataset for Multiple Instrument Recognition},'' in \emph{ISMIR 2018}, 2018.

\bibitem{fsdnoisy18k}
E.~Fonseca, M.~Plakal, D.~P.~W. Ellis, F.~Font, X.~Favory, and X.~Serra, ``Learning sound event classifiers from web audio with noisy labels,'' in \emph{ICASSP 2019}, 2019, pp. 21--25.

\bibitem{demand}
J.~Thiemann, N.~Ito, and E.~Vincent, ``{The Diverse Environments Multi-channel Acoustic Noise Database (DEMAND): A database of multichannel environmental noise recordings},'' in \emph{{21st International Congress on Acoustics}}, Jun. 2013.

\bibitem{adam}
D.~P. Kingma and J.~Ba, ``Adam: {A} method for stochastic optimization,'' in \emph{{ICLR} 2015}, May 2015.

\bibitem{dataaug}
A.~Alex, L.~Wang, P.~Gastaldo, and A.~Cavallaro, ``Data augmentation for speech separation,'' \emph{Speech Communication}, vol. 152, p. 102949, 2023.

\bibitem{dnsmos}
C.~K.~A. Reddy, V.~Gopal, and R.~Cutler, ``Dnsmos: A non-intrusive perceptual objective speech quality metric to evaluate noise suppressors,'' in \emph{ICASSP 2021}, 2021, pp. 6493--6497.

\bibitem{sedstats}
E.~Cakir, G.~Parascandolo, T.~Heittola, H.~Huttunen, T.~Virtanen, E.~Cakir, G.~Parascandolo, T.~Heittola, H.~Huttunen, and T.~Virtanen, ``Convolutional recurrent neural networks for polyphonic sound event detection,'' \emph{IEEE/ACM Trans. Audio, Speech and Lang. Proc.}, vol.~25, no.~6, p. 1291–1303, Jun. 2017.

\bibitem{mlaad}
N.~M. Müller, P.~Kawa, W.~H. Choong, E.~Casanova, E.~Gölge, T.~Müller, P.~Syga, P.~Sperl, and K.~Böttinger, ``Mlaad: The multi-language audio anti-spoofing dataset,'' in \emph{IJCNN 2024}, 2024, pp. 1--7.

\end{thebibliography}

\end{document}